\documentstyle[pre,twocolumn,aps,epsf,floats]{revtex}

\font\tenmsb=msbm10
\def\Bbb#1{\hbox{\tenmsb#1}}

\begin{document}
\draft
\title{Pattern selection in a lattice of pulse-coupled
 oscillators}

\author{X. Guardiola\cite{xaviermail} and
A. D\'{\i}az-Guilera\cite{albertmail}}
\address{
Departament de F\'{\i}sica Fonamental, Universitat de
 Barcelona, Diagonal 647, E-08028
Barcelona, Spain \\}

\date{\today}
\maketitle

\begin{abstract}
We study spatio-temporal pattern formation in a ring of
 $N$ oscillators with inhibitory
unidirectional pulselike interactions. The attractors of
 the dynamics are limit cycles
where each oscillator fires once and only once. Since
 some of these limit cycles lead
tothe same pattern, we introduce the concept of pattern
 degeneracy to take it into
account. Moreover, we give a qualitative estimation of
the volume of the basin of
attraction of each pattern by means of some probabilistic
 arguments and pattern
degeneracy, and show how are they modified as we change
 the value of the coupling
strength. In the limit of small coupling, our estimative
formula gives a perfect
agreement with numerical simulations.

\end{abstract}


\section{Introduction}
The study of the collective behavior of populations of
 interacting nonlinear oscillators
has attracted the interest of physicists and mathematicians
 for many years since they
can be used to modelize several chemical, biological and
 physical
systems\cite{Kura84,Win}. Among them, we should mention
cardiac pacemakers
cells\cite{Pes}, integrate and fire neurons\cite{Kura91}
 and other systems made of
excitable units\cite{Treves}. Most of the theoretical
 papers that have appeared in the
scientific literature deal with oscillators
 interacting through continuous-time
couplings, allowing them to describe the system by means
 of coupled differential
equations and apply most of the modern nonlinear
dynamics techniques. More challenging
from a theoretical point of view is to consider a
 pulse-coupling, or in other words,
oscillators coupled through instantaneous interactions
 that take place in a very
specific moment of its period. The richness of
behavior of these pulse-coupled
oscillatory systems includes synchronization
phenomena\cite{Mir}, spatio-temporal
pattern formation\cite{PhysD} (we could mention,
 for instance, traveling
waves\cite{Bres2}, chessboard structures\cite{PhysD},
 and periodic waves\cite{Rit} ),
rhythm anihilation\cite{nature}, self-organized
 criticality\cite{PRL1},...

Most of the work on pattern formation has been done
 in mean-field models or populations
of just a few oscillators. However, such restrictions
 do not allow to consider the
effect of certain variables whose effect can be crucial
 for realistic systems. The
specific topology of the connections or geometry of
 the system are some typical examples
which usually induce important changes in the collective
 behavior of these models.
Pattern formation usually takes place when oscillatory
 units interact in an inhibitory
way, although it has also been shown that the shape
 of the interacting pulse, when the
spike lasts for a certain amount of time, or time
 delays in the interactions can lead to
spatio-temporal pattern formation also in the case
 of excitatory
couplings\cite{Abbot,Geisel}. Only recently, general
 solutions for the general case,
where the patterns existence and stability is proved,
 have been worked
out\cite{Bres1,PRE}. The aim of this paper is to study
 some pattern properties and get a
quantitative estimation of the probability of pattern
 selection under arbitrary initial
conditions or, in the language of dynamical systems,
 the volume of the basin of
attraction of each pattern. Keeping this goal in mind,
 we will use the general results
given in \cite{PRE} where assuming a system defined on
 a ring the authors developed a
mathematical formalism powerful enough to get analytic
 information of the system. Not
only about the mechanisms which are responsible for
 synchronization and formation of
spatio-temporal structures, but also, as a complement,
 to proof under which conditions
they are stable solutions of the dynamical equations.

Despite the apparent simplicity of the model, some ring
 lattices of pulse-coupled
oscillators are currently used to modelize certain types
 of cardiac arhythmias where
there is an abnormally rapid heartbeat whose period is
 set by the time that an
excitation takes to travel the circuit \cite{Ito}. Moreover,
 there are experiments where
rings of a few R15 neurons from {\em Aplysia} are
 constructed and stable patterns are
reported \cite{Dror}. Our 1d model allows us to study
 analytically the most simple
patterns and understand their mechanisms of selection.

The structure of this paper is as follows. In Sec II we
 review the model introduced in
\cite{PRE} as well as set the notation used throughout
 the paper. In Section III we
study some pattern properties which will be useful for,
 in Section IV, propose an
estimation of the probability of selection of each
 pattern. In the last section we
present our conclusions.

\section{The model}

Our system consists in a ring of $(N+1)$ pulse-coupled
 oscillators. The phase of each
oscillator $\phi_i$ evolves linearly in time

\begin{equation}
\frac{d\phi_i}{dt}=1 \hspace{2em}\forall i=0,\ldots ,N
\end{equation}
until one of them reaches the threshold value $\phi_{th}=1$.
 When this happens the
oscillator fires and changes the state of its rightmost
 neighbor according to

\[
\phi_{i}\geq 1 \Rightarrow \left\{
\begin{array}{l}
\phi_{i}\rightarrow 0 \\ \phi_{i+1}\rightarrow\phi_{i+1}+
\varepsilon\phi_{i+1}\equiv\mu\phi_{i+1}
\end{array}
\right.
\]
subjected to periodic boundary conditions, i.e. $N+1\equiv 0$,
 and where $\varepsilon $
denotes the strength of the coupling and $\mu=1+\varepsilon$.
 Where we have assumed
that, from an effective point of view, the pulse-interaction
 between oscillators, as
well as the state of each unit of the system, can be described
 in terms of changes in
the phase, or in other words, in terms of the so called phase
 response curve (PRC),
$\varepsilon \phi$ in our case. A PRC for a given oscillator
 represents the phase
advance or delay as a result of receiving an external stimuli
 (the pulse) at different
moments in the cycle of the oscillator. We will assume
 $\varepsilon<0$ througout the
paper, as we are only interested in spatio-temporal
 pattern formation and
$\varepsilon>0$ always leads to the globally synchronized
 state\cite{PRE}. This linear
PRC has physical sense in some situations. For instance,
 it shows up when we expand the
non-linear PRC for the Peskin model of pacemaker cardiac
 cells \cite{Pes} in powers of
the convexity of the driving or in neuronal modelling\cite{Torras}.
 In practice,
however, this condition can be relaxed since a nonlinear
 PRC does not change the
qualitative behavior of the model provided the number of
 fixed points of the dynamics is
not altered. Moreover, a linear PRC has the advantage of
 making the system tractable
from an analytical point of view.

Let us describe the notation used in the paper. The population
 is ordered according to
the following criterion: The oscillator which fires will
 be always labeled as unit 0 and
the rest of the population will be ordered from this unit
 clockwise. After the firing,
the system is driven until another oscillator reaches the
 threshold. Then, we relabel
the units such that the oscillator at $\phi =1$ is now unit
 number 0, and so on. This
firing + driving (FD) process for $N+1$ oscillators can be
 described through a suitable
transformation

\begin{equation}
\vec{\phi}'=T_{k}(\vec{\phi})\equiv\vec{1}+{\Bbb{M}}_k\vec{\phi},
\end{equation}
where ${\Bbb{M}}_k$ is a $N \times N$ matrix, $\vec{\phi}$
 is a vector with $N$
components, $\vec{1}$ is a vector with all its components
 equal to one and $k$ stands
for the index of the oscillator which will fire next.
 We call this kind of
transformation a firing map, and we have to define
 as many firing maps as oscillators
could fire, that is, index $k$ must run from $k=1$
($\phi_1$ fires) to $N$ ($\phi_N$
fires). For example, in the $N+1=4$ oscillators
 case we have that the firing map
corresponding to the FD process where $\phi_2$ is
the next oscillator which do fire,

\vspace{5mm}

\begin{tabular}{ccccc}
$\phi_0=1$ & $\stackrel{\mbox{firing}}{\longrightarrow}$
 & $0$ &
 $\stackrel{\mbox{driving}}{\longrightarrow}$ &
 $1-\phi_2=\phi'_2 $ \\
$\phi_1$ & $\longrightarrow$ & $\mu\phi_1$ &
 $\longrightarrow$ &
 $\mu\phi_1+ 1-\phi_2=\phi'_3$ \\
$\phi_2$ & $\longrightarrow$ & $\phi_2$ & $\longrightarrow$
& $1=\phi'_0$ \\ $\phi_3$ &
$\longrightarrow$ & $\phi_3$ & $\longrightarrow$ &
 $\phi_3+ 1-\phi_2=\phi'_1$
\end{tabular}
\vspace{5mm}

\noindent would be $T_2(\vec{\phi})$

\begin{equation}
\left(
\begin{array}{c}
\phi^{\prime}_1 \\ \phi^{\prime}_2 \\ \phi^{\prime}_3
\end{array}
\right) = \left(
\begin{array}{c}
1 \\ 1 \\ 1
\end{array}
\right) + \underbrace{\left(
\begin{array}{ccc}
0 & -1 & 1 \\ 0 & -1 & 0 \\ -\mu & - 1 & 0
\end{array}
\right) }_{{\Bbb{M}}_2} \left(
\begin{array}{c}
\phi_1 \\ \phi_2 \\ \phi_3
\end{array}
\right)
\end{equation}
and so on. Once we have defined all possible firing
maps for a given number of
oscillators we can proceed to deal with the attractors
 or fixed points of the system
dynamics. As has been proved in \cite{PRE} these fixed
 points must be cycles of $N+1$
firings. We define a cycle as a sequence of consecutive
 firings where each oscillator
fires once and only once. Mathematically, each cycle is
 described by means of a return
map. The return map is the transformation that gives the
evolution of $\vec{\phi}$
during a cycle and is the composition of all firing maps
 involved in the firing sequence
of that cycle

\begin{equation}
\vec{\phi}'=T_{c_1} \circ T_{c_2}\ldots \circ T_{c_{N+1}}
(\vec{\phi})\equiv\
\vec{R}_{c}+{\Bbb{M}}_{c}\vec{\phi},
\end{equation}
where $T_{c_i} \circ T_{c_j}(\phi)$ is the usual
composition operation
$T_{c_i}(T_{c_j}(\phi))$ and

\[
\vec{R}_{c}=\vec{1}+\sum_{i=c_1}^{c_N}(\prod_{j=c_1}^{i}
{\Bbb{M}}_{j}) \cdot\vec{1}
\hspace{2em} \mbox{and} \hspace{2em} {\Bbb{M}}_{c}=
\prod_{j=c_1}^{c_{N+1}}{\Bbb{M}}_{j}.
\]
Note that not all possible combinations of firing maps
 are allowed, just those ones
whose indices $c_i$ sum $p(N+1)$ without any partial
 sum equal to $q(N+1)$, where $p>q$
are positive integers.

As all firing maps are linear transformations, return maps
 are also linear. There are
$N!$ possible cycles in the $N+1$ oscillators case (all
 permutations of firing sequences
with the initial firing oscillator $\phi_0$ fixed). Following
 our previous example, for
the four oscillators case all possible firing sequences
 and their associated return maps
are

\begin{eqnarray*}
A: 0,1,2,3 \rightarrow T_1 \circ T_1 \circ T_1 \circ T_1
\\ B: 0,1,3,2 \rightarrow T_2
\circ T_{3} \circ T_{2} \circ T_{1} \\ C: 0,2,1,3 \rightarrow
 T_{1} \circ T_{2} \circ
T_{3} \circ T_{2} \\ D: 0,2,3,1 \rightarrow T_{3} \circ
T_{2} \circ T_{1} \circ T_{2}
\\ E: 0,3,1,2 \rightarrow T_{2} \circ T_{1} \circ T_{2}
 \circ T_{3} \\ F: 0,3,2,1
\rightarrow T_{3} \circ T_{3} \circ T_{3} \circ T_{3}
\end{eqnarray*}
Now, in order to find the attractors of the dynamics,
we must solve the fixed point
equation

\begin{equation}
\vec{\phi}_{c}^{*}=\vec{R}_{c}+\Bbb{M}_{c}\vec{\phi}_{c}^{*},
\end{equation}
for every cycle $c$. Formally,

\begin{equation}
\vec{\phi}_{c}^{*}=\vec{R}_{c}\cdot({\Bbb{I}}-{\Bbb{M}}_{c})^{-1}.
\end{equation}
As was shown in \cite{PRE}, there are $N$ different stable
 solutions to the whole set of
fixed point equations. Their stability is assured by the
 fact that $\varepsilon<0$,
since it guarantees that all eigenvalues of $\Bbb{M}_{c}$
 lie inside the unit circle for
all cycles $c$. In our four oscillators example these
 solutions are

\begin{eqnarray*}
\nonumber \vec{\phi}_{A}^{*}= & (1,\frac{3}{4+3\varepsilon},
 \frac{2}{4+3\varepsilon},
\frac{1}{4+3\varepsilon}) & \hspace{1em}  \\ \nonumber
\vec{\phi}_{B}^{*}=\vec{\phi}_{C}^{*}=\vec{\phi}_{D}^{*}
 =\vec{\phi}_{E}^{*}= &
(1,\frac{1} {2+\varepsilon},1,\frac{1}{2+\varepsilon})
 & \hspace{1em}  \\
\vec{\phi}_{F}^{*}= & (1,\frac{1}{4+\varepsilon},
\frac{2+\varepsilon}{4+
\varepsilon},\frac{3+\varepsilon}{4+\varepsilon}) &
 \hspace{1em}
\end{eqnarray*}
Which are a kind of four-oscillators traveling wave,
chessboard and inverse traveling
wave structures.

From now on we will label such solutions with index $m$
($m=1...N$) since their first
component always satisfy

\begin{equation}
\phi_{1}^{*}=\frac{m}{N+1+m\varepsilon}.
\end{equation}
Therefore, in the example, we relabel patterns
$\vec{\phi}_{A}^{*}$ as $m=3$,
$\vec{\phi}_{B}^{*}$, $\vec{\phi}_{C}^{*}$,
$\vec{\phi}_{D}^{*}$, $\vec{\phi}_{E}^{*}$
as $m=2$ and $\vec{\phi}_{F}^{*}$ as $m=1$.

Since there are $N!$ possible cycles and $N$ solutions
 to Eq. (7) there will be some
fixed points or patterns which will appear more than
 once, so, we shall use $C(N+1,m)$
to characterize these degeneracies. In the example,
the values of the degeneracies are
$C(4,1)=C(4,3)=1$ and $C(4,2)=4$. In general, patterns
which are solutions of cycle
consisting in the iterative application of the same
firing map (like A and F in our
example) have no periodicities whereas the ones
solution of mixtures of differents
firing maps (B,C,D and E) have some periodic structure
 that are also solution of Eq. (7)
for a case with less oscillators. In Fig. 1 we can
 visualize the solutions for $N+1=2,3$
and $4$ oscillators and realize that solution $m=2$
for the four oscillators case is a
periodic composition of solution $m=1$ for the two
oscillators case.

\begin{figure}
\centerline{
        \epsfxsize= 8.0cm
        \epsffile{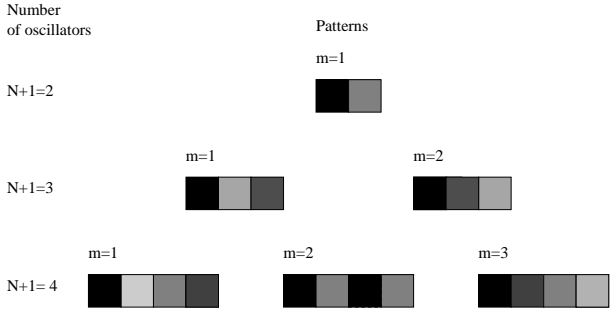}}
        \caption{Graphic representation of the patterns solution of Eq. (7)
        for a small value of the coupling strength $|\varepsilon|$
        at the beginning of the cycle (we must keep in mind that
        spatio-temporal patterns
        are dynamical structures that evolve in time).
        The leftmost square represents $\phi_0$ and the rightmost $\phi_N$,
        and their phase is visualized in a greyscale where black means
        $\phi=1$ and white $\phi=0$. }
\end{figure}

\section{Pattern properties}

As we have seen, the stability of all patterns solution
of Eq. (6) is guaranteed by the
fact that $\varepsilon<0$, but the existence of such
 solutions is not ensured. In fact,
for small values of the coupling strength $|\varepsilon|$
 all patterns do exist, but, as
we increase it, some patterns disappear. The reason is
 that the solution loses its
physical meaning because $\phi_1^*>1$. Their first component
 is always the one that
becomes larger than unity earlier and this happens, for
each $m$ and according to Eq.
(9), when

\begin{equation}
\varepsilon<\varepsilon_m^*=1-\frac{N+1}{m}.
\end{equation}
Our coupling strength range of interest ends at
$\varepsilon=-1$, since at
$\varepsilon\leq -1$ we always find the same pathological
 dynamics which does not have
any physical or biological sense. Realistic couplings
never reach such higher values.
Therefore, as $\varepsilon$ runs from $0$ to $-1$, all
 patterns whose $m$ satisfy
$m>\frac{N+1}{2}$, disappear.

There is another interesting pattern property which has
 to do with the calculation of
the pattern degeneracy $C(N+1,m)$. In principle, to
 calculate such degeneration, we
should solve fixed point Eq. (6) for all possible cycles
 and count how many of them lead
to the same pattern. Although for few oscillators the
problem is quite straightforward,
as we deal with higher and higher number of oscillators,
the number of cycles increases
(it grows as $N!$) and solving Eq. (6) becomes more difficult.
 Fortunately, there is
another way of calculating $C(N+1,m)$ which reduces
the problem to a combinatorial
question. Lets show it through an example, in the
previous four oscillators case, if we
count, for each firing sequence, the number of oscillators
 which have received the pulse
before firing, we can easily realize that this number
 is the same as its value of $m$

\begin{eqnarray*}
A: & 0,\overline{1},\overline{2},\overline{3} &
\hspace{2em} m=3 \\ B: &
0,\overline{1},3,\overline{2} & \hspace{2em} m=2 \\ C:
& 0,2,\overline{1},\overline{3} &
\hspace{2em} m=2 \\ D: & 0,2,\overline{3},\overline{1}
& \hspace{2em} m=2 \\ E: &
0,3,\overline{1},\overline{2} & \hspace{2em} m=2 \\ F:
& 0,3,2,\overline{1} &
\hspace{2em} m=1
\end{eqnarray*}
Here an upper bar means that the oscillator has already
 received a pulse during the
cycle. The point is that it turns out that every pattern
 $m$ corresponds to a sequences
of firings involving exactly $m$ oscillators that, when
 they do fire, had already
received a pulse from their leftmost neighbor. Therefore,
 this property (we have checked
for several values of $N+1$) allows us to associate every
 cycle with the pattern it
leads to, just by counting these kind of firings. Now,
 calculating $C(N+1,m)$ becomes a
straightforward matter. In Table I we have computed
$C(N+1,m)$ for several values of
$N+1$.

\begin{table}
\caption{Pattern degeneracy $C(N+1,m)$. First column
 stands for the number $N+1$ of
oscillators and first row for $m$.}
\begin{tabular}{cccccccccc}
& 1 & 2 & 3 & 4 & 5 & 6 & 7 & 8 & 9\\ \hline 2 & 1 &
  &  &  &  \\ 3 & 1 & 1 &  &  &  \\
4 & 1 & 4 & 1 &  &  \\ 5 & 1 & 11 & 11 & 1 &  \\ 6 &
 1 & 26 & 66 & 26 & 1 \\ 7 & 1 & 57
& 302 & 302 & 57 & 1 \\ 8 & 1 & 120 & 1191 & 2416 &
 1191 & 120 & 1 \\ 9 & 1 & 247 & 4293
& 15619 & 15619 & 4293 & 247 & 1 \\ 10 & 1 & 502 &
 14608 & 88234 & 156190 & 88234 &
14608 & 502 & 1
\end{tabular}
\end{table}
Apart from brute force counting, degeneracy distribution
 $C(N+1,m)$ can also be
determined from the following relation

\begin{eqnarray}
\nonumber C(N+1,m)& = & mC(N,m)+ \\ & &(N+1-m)C(N,m-1),
\end{eqnarray}
for $2\leq m\leq N-1$. This recursion relation is closed by

\begin{equation} C(N+1,1)=C(N+1,N)=1,
\end{equation}
which correspond to the firing sequences

\[
0,N,(N-1)...2,\overline{1} \hspace{2em} \mbox{and} \hspace{2em}
0,\overline{1},\overline{2}...\overline{(N-1)},\overline{N},
\]
respectively.

From the previous relations one can deduce by induction
 the symmetry of the distribution
with respect to its extremes at $m=1$ and $m=N$

\begin{equation}
C(N+1,m)=C(N+1,N+1-m),
\end{equation}
and

\begin{equation}
\sum_{m}C(N+1,m)=N!.
\end{equation}
Another interesting property is the period $\Delta_{m}^{N+1}$
 of each spatio-temporal
pattern $m$. Since all oscillators are in a phase-locked
state, they must oscillate with
the same period. Then, as the intrinsic period of each
 oscillator is one, and when any
oscillator receives the delaying pulse from its
neighbor it has a phase equal to
$\phi_{1}^{*}$, one can easily realize that the
effective period is

\begin{equation}
\Delta_{m}^{N+1}=1+\varepsilon \phi_{1}^{*}=
\frac{N+1+2m\varepsilon}{N+ 1+m\varepsilon}.
\end{equation}
Therefore, the larger the value of $m$, the longer the period
 of its associated pattern.
It is important to notice that we have not fixed the
value of such periods (each pattern
has its own which is different from the others),
since there are some authors who fix
all periods equal to some constant, and use
it as a condition to find the
structures\cite{Dror}.

\section{Pattern selection}

Once we have characterized all spatio-temporal patterns,
we proceed to find some general
formula which give us some estimation of the probability
 of each pattern to be selected,
or in other words, an estimation of the volume of its
 basin of attraction. In order to
achieve this objective, we should understand the mechanism
which lead to the selection
of a certain spatio-temporal structure and how is it
 modified as the parameters of the
model ($\varepsilon$ in our case) change.

There is an easy and straightforward way to get the
 essential features of this mechanism
assuming that the probability of one oscillator to fire
next is, basically, proportional
to its phase (that is, if it has a phase slightly below
$1$ it has a higher probability
to be the next firing oscillator, whereas if it has a
smaller phase, it will rarely fire
next). Imagine the phases of all oscillators randomly
 distributed over the interval
$(0,1)$. Then we let the system evolve till one of the
 oscillators reaches a phase
$\phi_i=1$ and emits a pulse that is received by its
 rightmost neighbor which lows its
phase by an amount $\varepsilon \phi_{i+1}$. Now we
 assume that all phases are again
randomly distributed over $(0,1)$ except the one which
 received the pulse whose phase is
distributed over $(0,1+\varepsilon)$. So, we get rid
of memory effects (we know the
oscillator that has fired should, now, have a phase
equal to zero) and just keep in mind
if each oscillator has received a pulse or has not.
 Therefore, the point is that under
this conditions, the probability that one oscillator
 which has still not received a
pulse do fire is some constant and, on the other hand,
for the ones which had, is this
constant times the factor $(1+\varepsilon)$. Then, we
 can characterize the probability
of having some cycle just by recalling how many
 oscillators do fire having previously
received a pulse during that cycle. Basically, this
 probability is proportional to
$(1+\varepsilon)^{n}$ where $n$ stands for the number
 of oscillators which do fire
having already received a pulse (the product of all
 constant terms will be absorbed in a
normalization factor). This approach, where we assume
 all firings as almost-independent
events, can be viewed as a kind of mean-field
 approximation. Then,as has been shown
before, since cycles leading to the same pattern $m$
 always exactly have $m$ oscillators
that do fire having received the interacting pulse,
 we can give an estimation of the
probability for pattern $m$ selection in the $N+1$
 oscillators case

\begin{equation}
p_m^{N+1}(\varepsilon) \simeq {\cal N}(\varepsilon)C(N+1,m)
 (1+\varepsilon)^m.
\end{equation}
Here ${\cal N}(\varepsilon)$ is chosen so that summation
of the probabilities over m
gives 1

\begin{equation}
\sum_m p_m^{N+1}(\varepsilon)=1.
\end{equation}
In the limit of small coupling strength $\varepsilon\rightarrow 0$,
 which is the more
interesting case for the majority of physical and
 biological systems, one can assume
that interaction plays almost no role when pattern
 selection takes place. That is, the
fact that one oscillator has received the pulse from
its neighbor does not low its
probability to fire as the pulse does not modify
appreciably its phase. Then, we can
consider that all cycles have approximately the
 same probability to be selected,
$(1+\varepsilon)^m \rightarrow 1$, and only pattern
 degeneracy has to be considered to
get a good estimation of $p_m^{N+1}$

\begin{equation}
p_m^{N+1} \simeq \frac{C(N+1,m)}{N!}.
\end{equation}
The dominant pattern, that is, the one which has the
 larger probability to be selected
coincides with the mean value of $m$ (due to the
 symmetric behavior of $C(N+1,m)$).

\begin{equation}
<m>_{N+1}=\sum_{m}m\frac{C(N+1,m)}{N!}=\frac{N+1}{2}.
\end{equation}
For an odd number of oscillators $<m>_{N+1}$ does
not exist and we have a competition
between the two closest patterns $m=N/2$ and $m=(N+2)/2$.
 Recall that the most probable
patterns turn out to be the ones with "shortest wavelengths",
 a fact that was already
reported in simulations of these sort of systems\cite{PhysD}.
 In Figs. 2 and 3 we check
this new approximation for the $N+1=10$ and $9$ case
 and realize that expected results
are in good agreement with simulations data.

There also is the interesting question of how does
this probability distribution
modifies when the number of oscillators increases.
 In Fig. 4 we show $p_{m}$ for
different values of $N+1$. Since there are more
possible values of $m$ available, as we
increase $N+1$, $p^{N+1}_{m}$ diminishes. The distribution
also gets narrower as we
increase $N+1$ and this becomes clear when one
studies the variance of $p_{m}$. It can
be found that

\begin{eqnarray}
\nonumber <m^{2}>_{N+1} & = & \sum_{m}m^{2}
\frac{C(N+1,m)}{N!} \\ & =
&\frac{(N+1)^{2}}{4}+\frac{N+1}{12}.
\end{eqnarray}
We could not prove this without an explicit expression
 for $C(N+1,m)$ but we have
checked it N up to $170$. Therefore

\begin{equation}
\sigma^{2}_{N+1}=\frac{N+1}{12}=\frac{<m>_{N+1}}{6}.
\end{equation}
It turns out that for a large number of oscillators
 almost all initial conditions lead
to a pattern whose $m$ approximately falls in the
 interval $<m>_{N+1}\pm
\sqrt{<m>_{N+1}}$. In order to compare it for
 different number of oscillators we have to
normalize $m$ dividing by $N+1$. In that case,
 one observes that $\sigma^{2}_{N+1}\sim
1/\sqrt{N+1}$ so that as we increase $N+1$, the spread
 of $p^{N+1}_{m}$ diminishes
getting the distribution sharpened.

\begin{figure}
\centerline{
        \epsfxsize= 6.0cm
        \epsffile{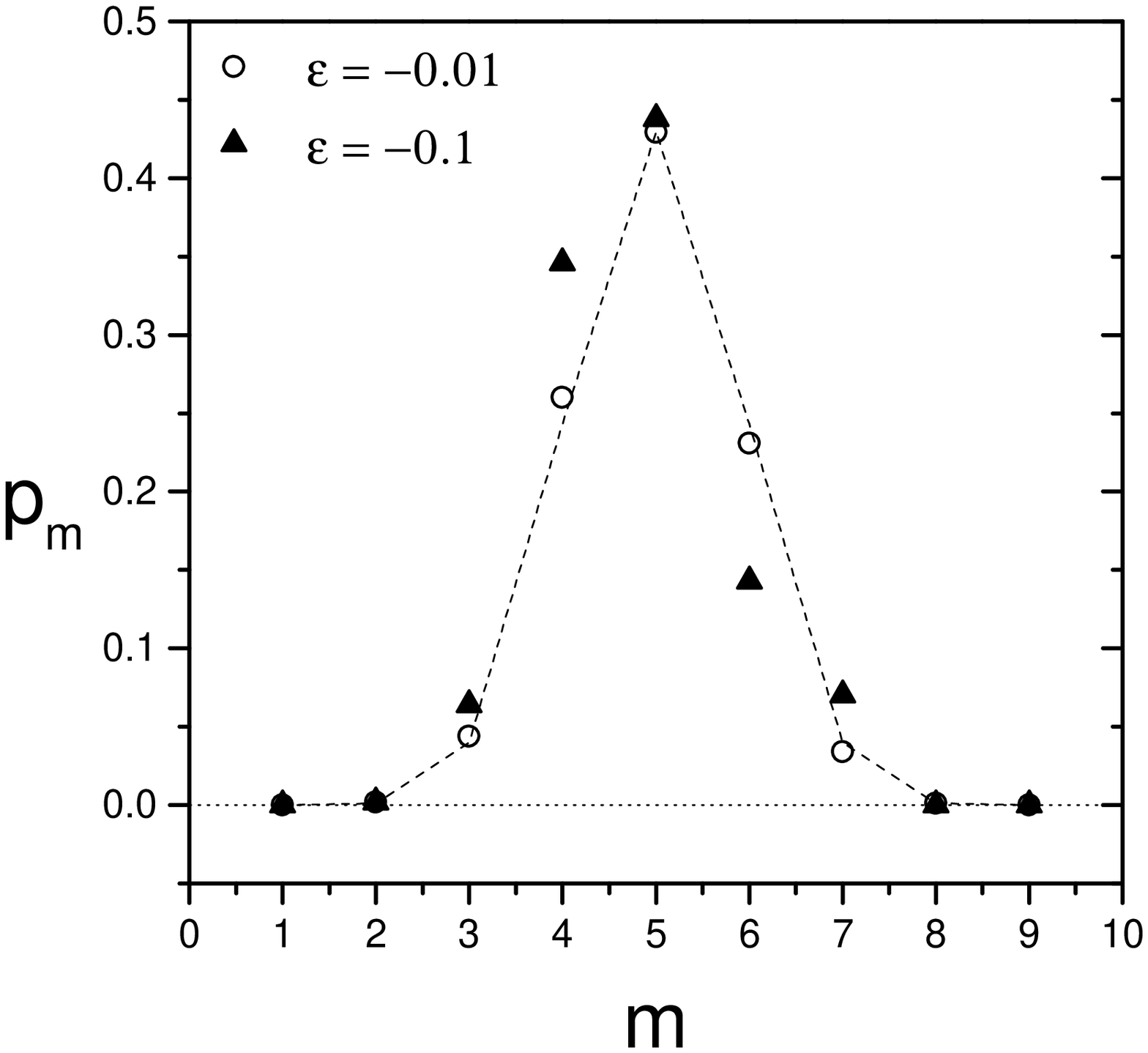}}
        \caption{Estimation of $p_m$ for small
        coupling strength in the case of an even number of oscillators
        $N+1=10$ and for a values of $\varepsilon$
        equal to $-0.01$ and $-0.1$. Dashed line follows the theoretical
        values estimated through Eq. (16). We can realize that the smaller
        $|\varepsilon|$ is, the more accurate our estimations are.
        The most probable pattern is $m=(N+1)/2$ and the
        probability for the patterns near the extremes is almost zero due
        to the fast decay of $p_m$ there.}
\end{figure}

\begin{figure}[t]
\centerline{
        \epsfxsize= 6.0cm
        \epsffile{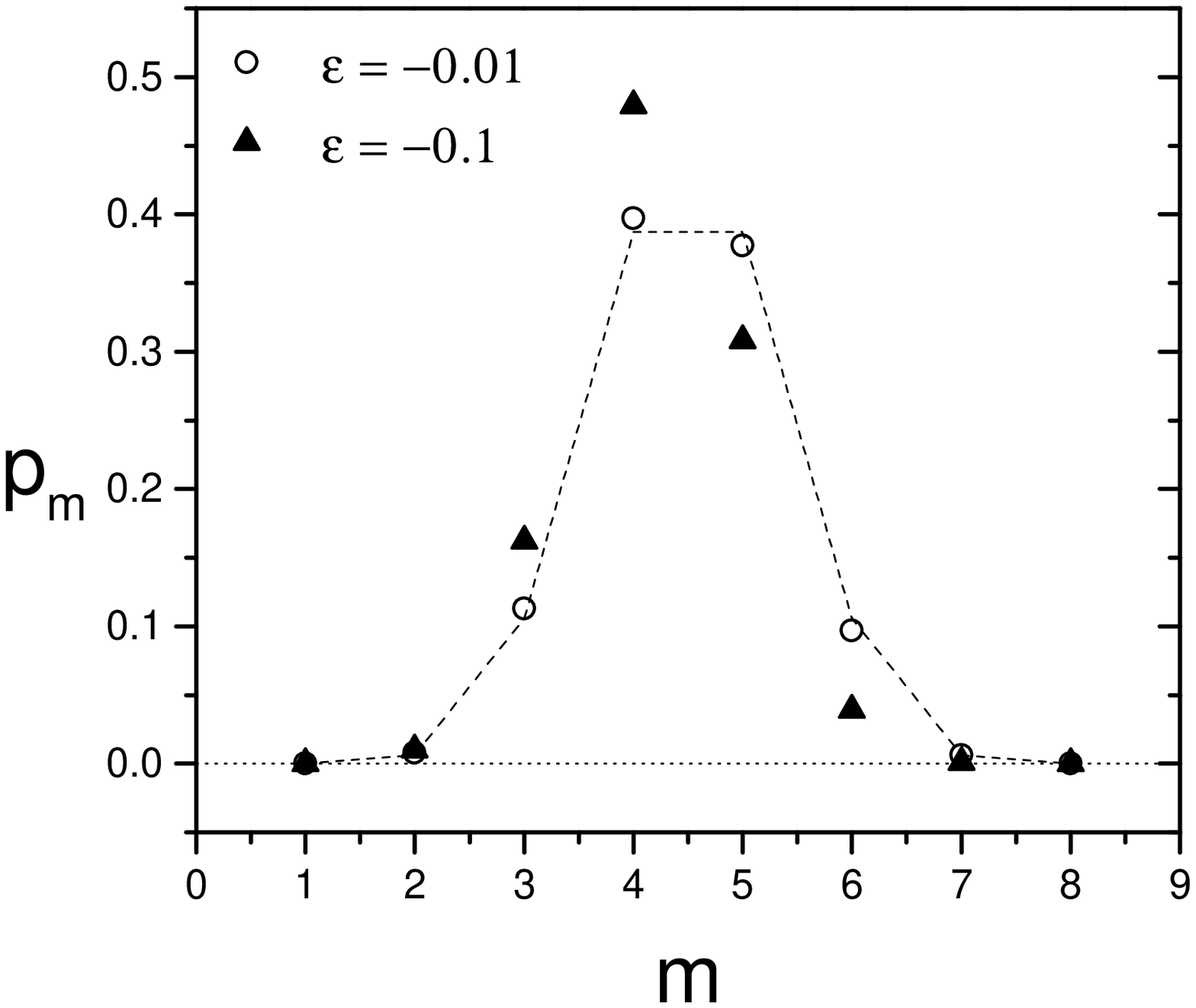}}
        \caption{The same as Fig. 2 but now for an odd number of oscillators
        $N+1=9$. We can realize that there is not a peak anymore, instead,
        almost all probability is concentrated in the two competing patterns
        $m=N/2$ and $m=(N+2)/2$.}
\end{figure}

\begin{figure}[t]
\centerline{
        \epsfxsize= 6.0cm
        \epsffile{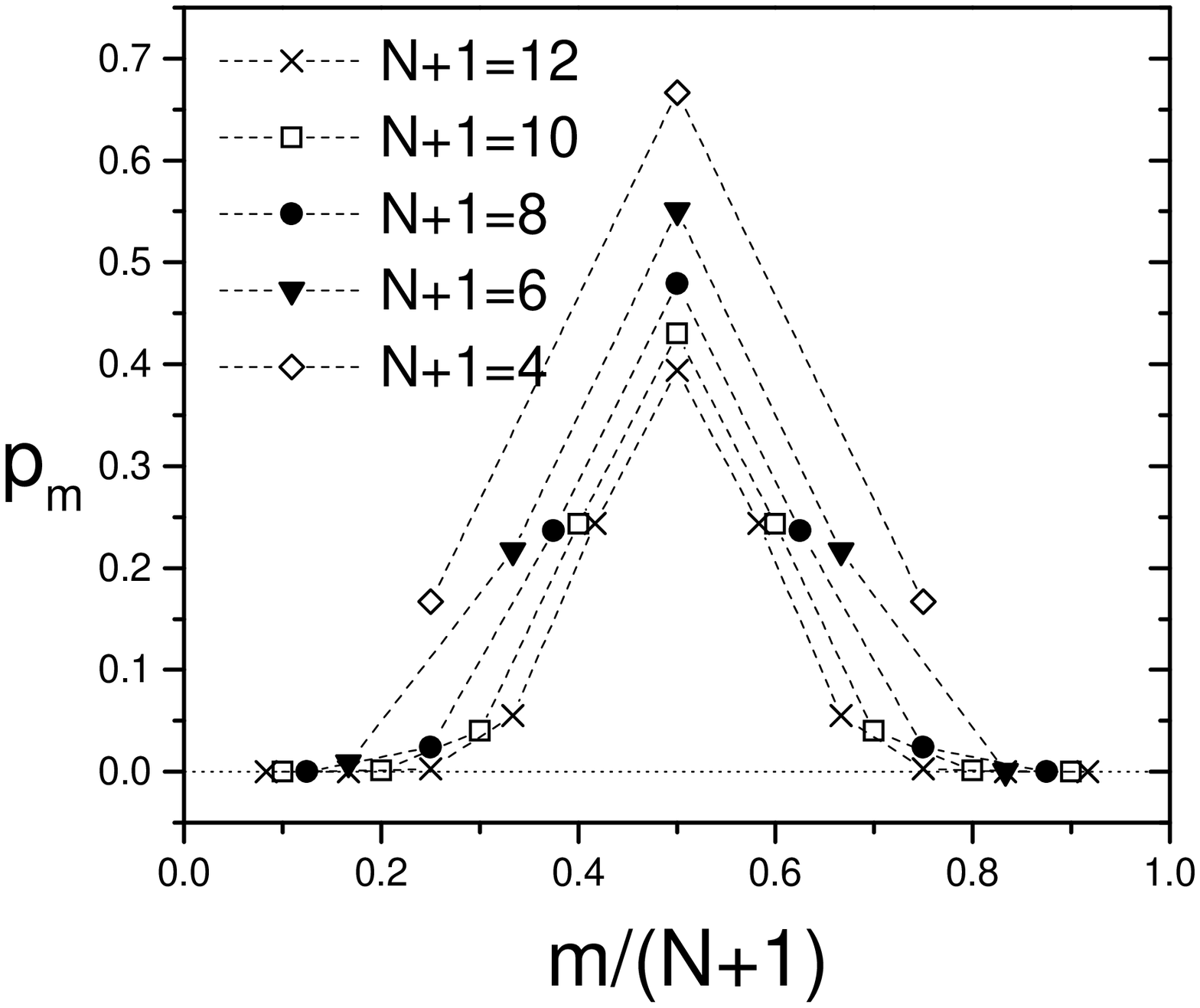}}
        \caption{$p_m$ calculated by means of Eq. (17) for different values of
        $N+1$. The distribution gets narrower and the
        height of its peak diminishes as we increase the number of oscillators.
                       }
\end{figure}

As Eq. (14) does not take into account the disappearance
 of the different patterns $m$
at the different values of $\varepsilon_{m}^{*}$
 predicted by Eq. (9), it can not give a
good quantitative estimation of pattern selection for
 higher coupling values.
Nevertheless we can expand Eq. (14) to the leading
 order in $\varepsilon$. For small
$\varepsilon$, $p_m^{N+1}$ are approximated by
\begin{equation}
p_m^{N+1}\simeq \frac{C(N+1,m)}{N!}(1+(m-\frac{N+1}{2})\varepsilon).
\end{equation}
In Fig. 5 we compare this approximation with simulated data.
 The slopes near
$\varepsilon=0$ do agree with Eq. (21).
\begin{figure}
\centerline{
        \epsfxsize= 6.0cm
        \epsffile{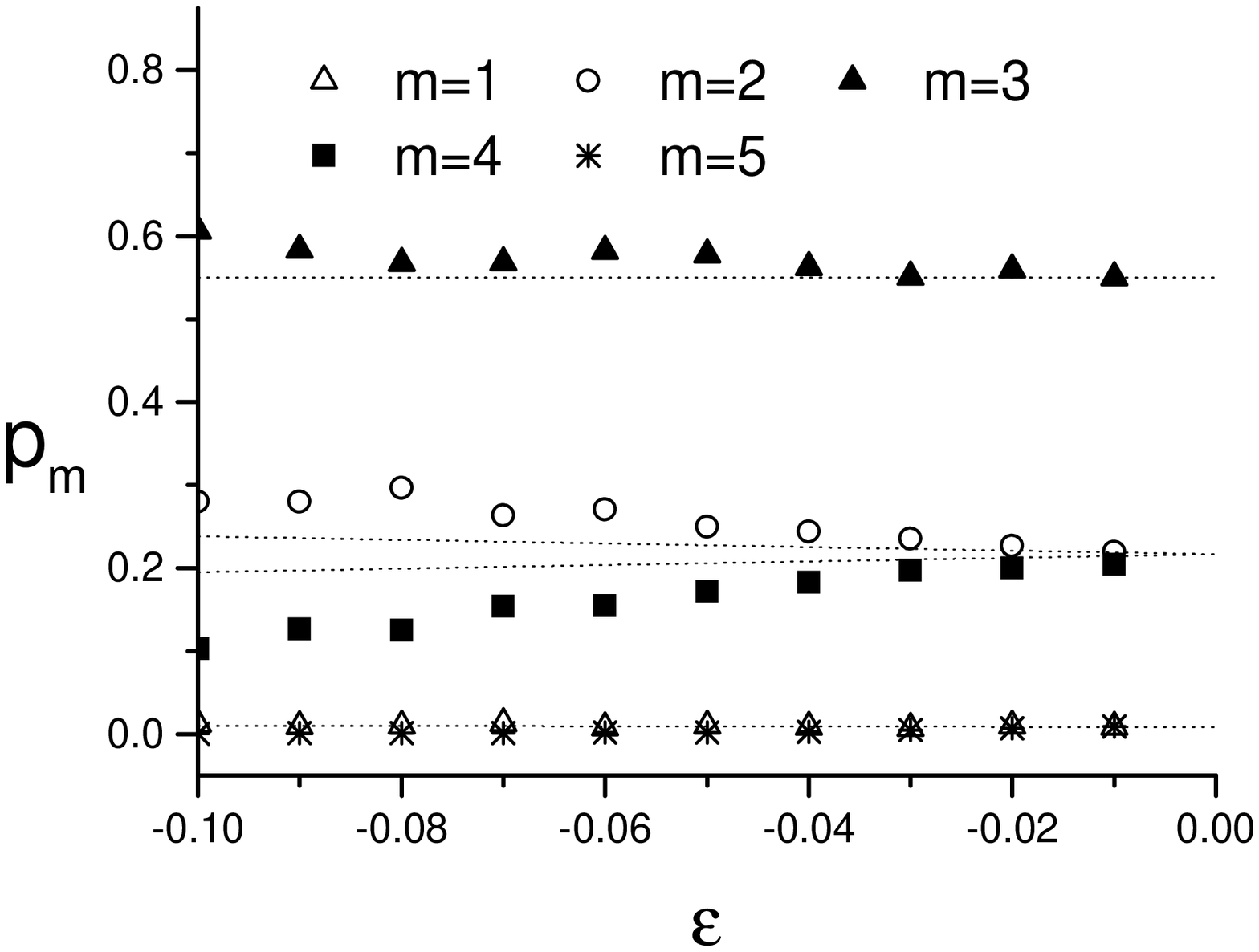}}
        \caption{Estimation of $p_m^{N+1}$ for a system of N+1=6 oscillators
        by means of the linearized Eq. (21).
        For small $\varepsilon$ the slopes match with simulated data. }
\end{figure}
In our simulations we calculate the probability of
 each pattern to be selected just by
counting how many realizations (with $\phi_0=1$ and
 the rest of oscillators with random
initial conditions) lead to each pattern $m$ and divide
 over the total number of
realizations.
\begin{figure}
\centerline{
        \epsfxsize= 6.0cm
        \epsffile{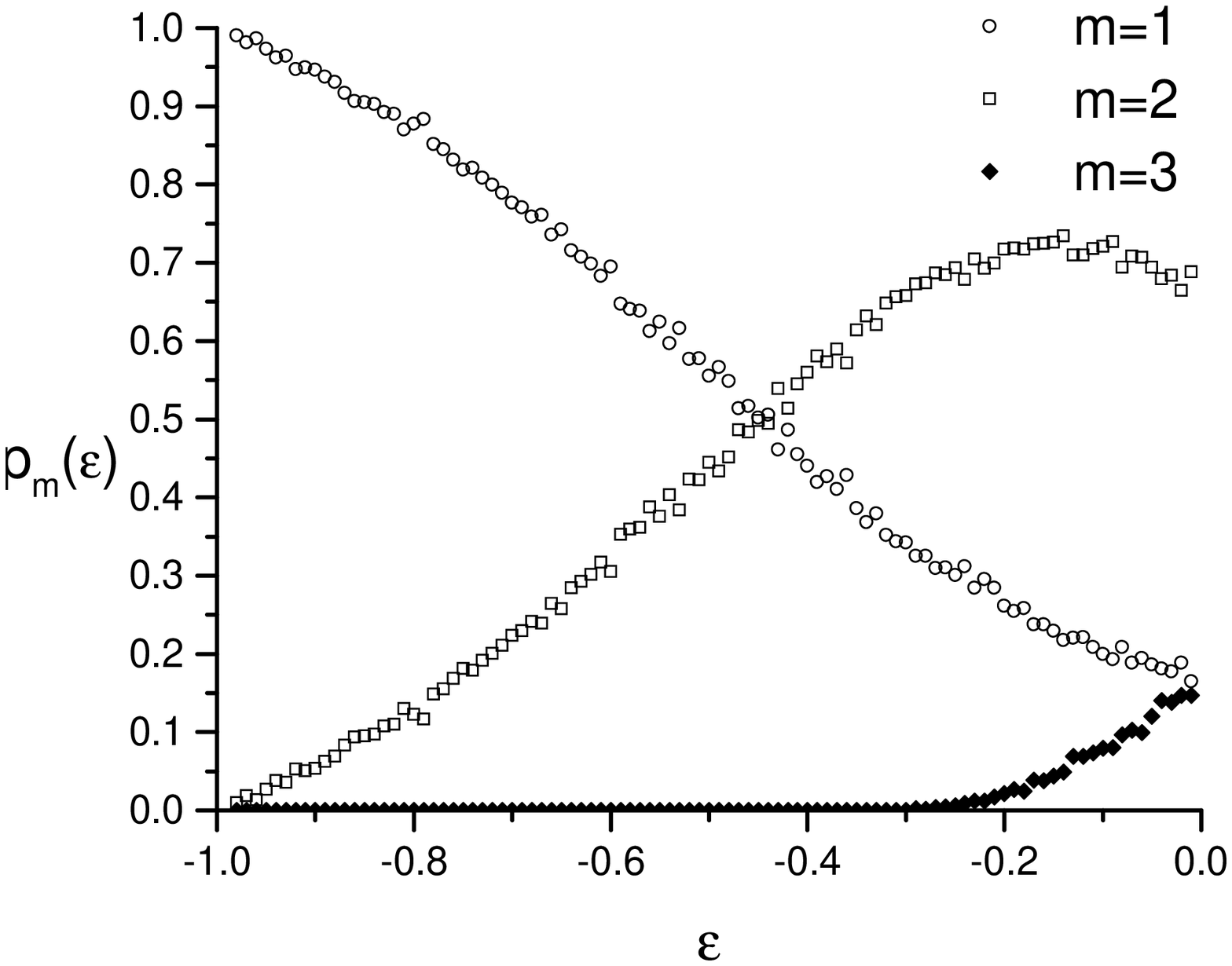}}
        \caption{$p_m(\varepsilon)$ in the case of $N+1=4$ oscillators.
        In this case, we see that
        although for small coupling chessboard ($m=2$) is the dominant
        pattern, the inverse traveling wave ($m=1$) is the most probable
        pattern for higher values of the coupling strength. Simulations
        are done over 2500 realizations.}
\end{figure}
\begin{figure}
\centerline{
        \epsfxsize= 6.0cm
        \epsffile{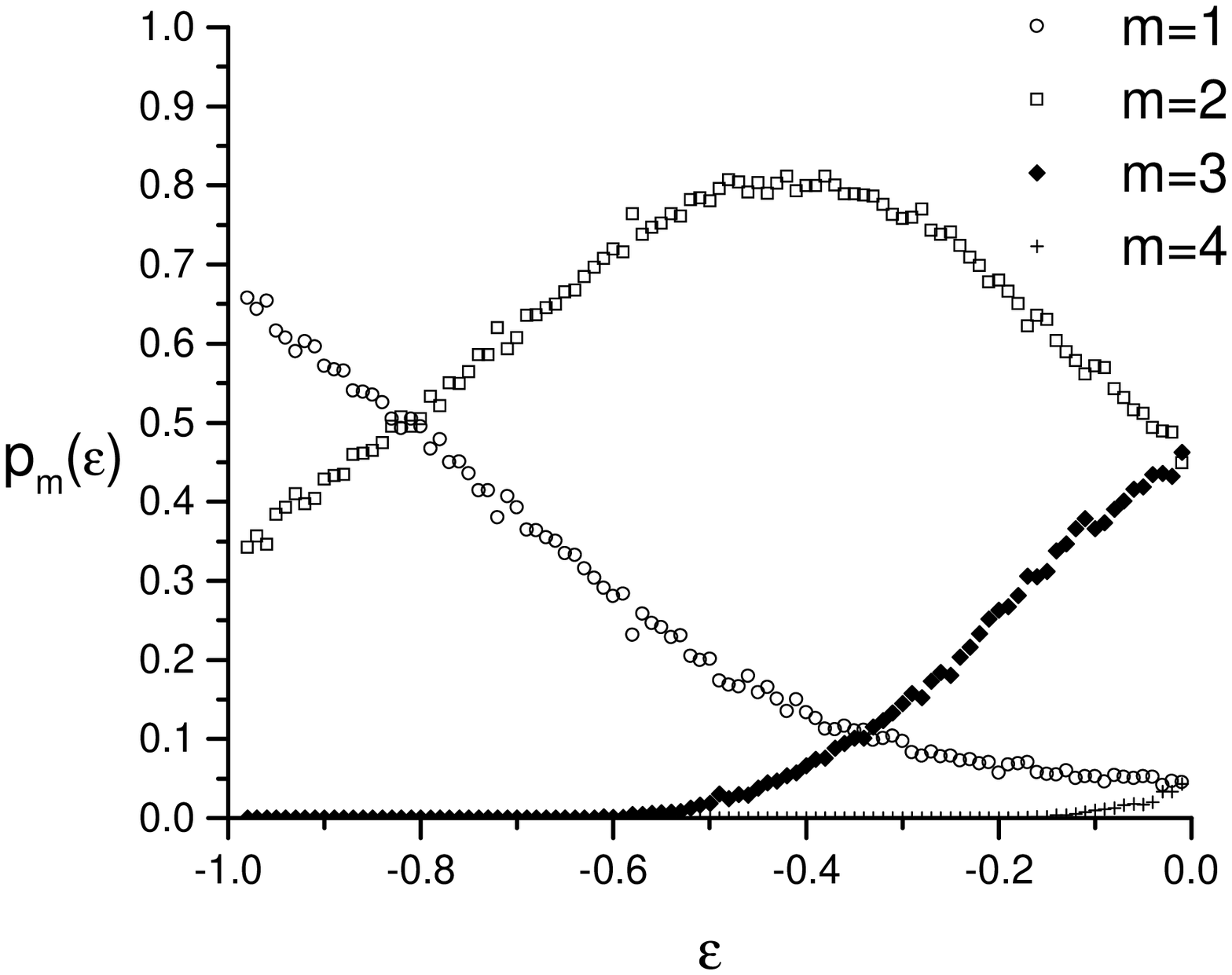}}
        \caption{The same as Fig. 6 for $N+1=5$.}
\end{figure}
\begin{figure}
\centerline{
        \epsfxsize= 6.0cm
        \epsffile{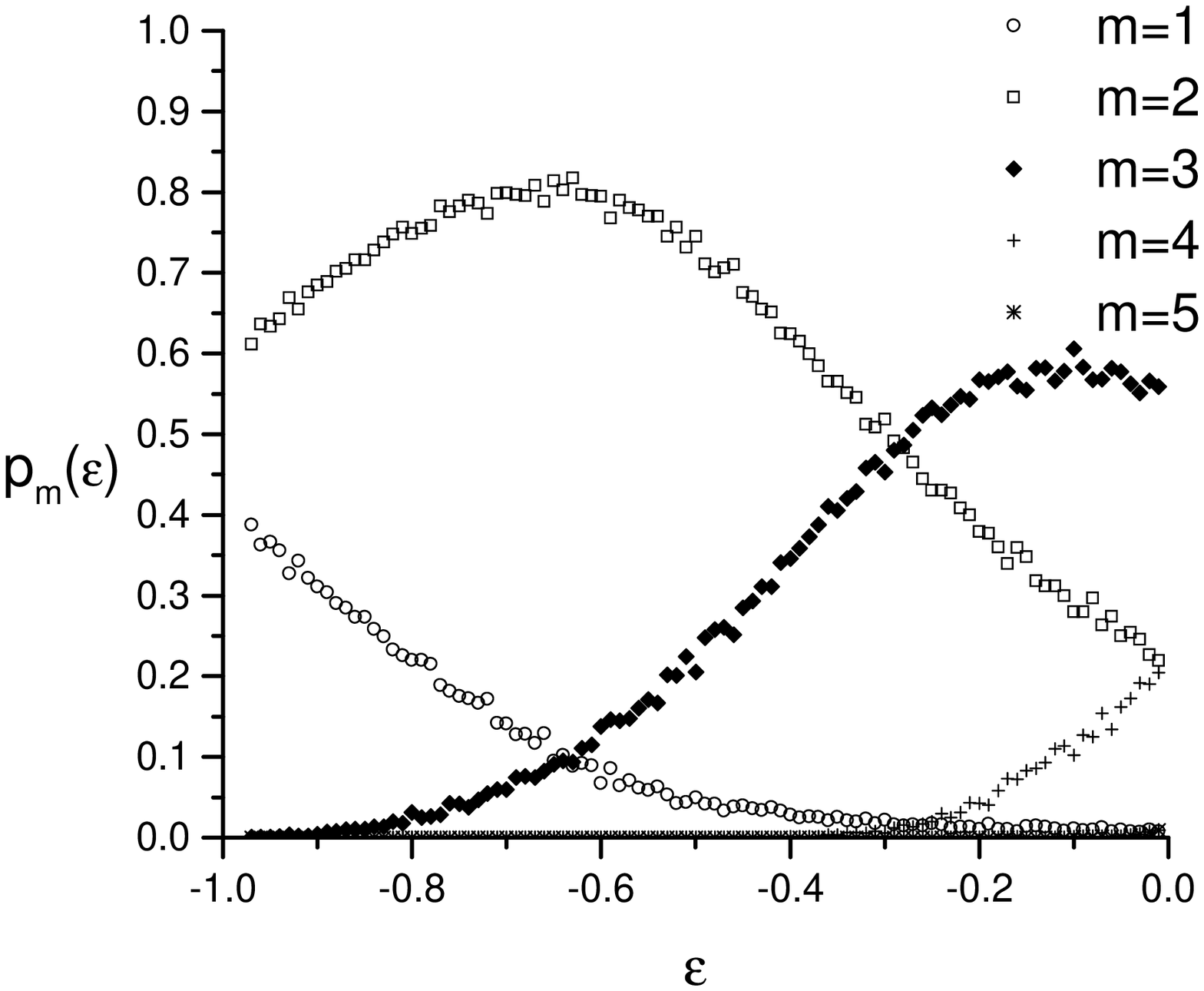}}
        \caption{The same as Fig. 6 for $N+1=6$}
\end{figure}
Although we only have a good quantitative estimation of
 $p^{N+1}_{m}$ for small values
of $\varepsilon$, Eq. (15) catches the two basic mechanisms
 responsible of pattern
selection. On the one hand, it is clear that for higher
 values of the coupling strength
$|\varepsilon|$, when one oscillator receives a pulse,
it lows its phase to almost zero
and, consequently, its firing probability also does.
Therefore pattern selection
probability $p_m^{N+1}(\varepsilon)$ is strongly controlled
 by the number of oscillators
which have to fire having already received a pulse, that
 is, the probabilistic factor
$(1+\varepsilon)^m$. As a consequence, $p^{N+1}_{m}$ begin
 to decrease sooner when
$|\varepsilon|$ increases, the larger $m$ is. On the other
 hand, for small values of the
coupling strength, interaction plays almost no role and
 $p_m^{N+1}(\varepsilon)$ is
dominated by the degeneracy factor $C(N+1,m)$. Therefore
$p_m^{N+1}(\varepsilon)$ for
the different values of $m$ are basically ordered as
$C(N+1,m)$. In Fig. $6$, $7$ and
$8$ we show results from simulations of $p_m^{N+1}(\varepsilon)$
for different number of
oscillators.

\section{Conclusions}

In this paper we have studied some properties of the
spatio-temporal patterns that
appear in a ring of pulse-coupled oscillators with
inhibitory interactions. We have
focused our attention in estimating the probability
of selecting a certain pattern under
arbitrary initial conditions and have shown the two
basic mechanisms responsible of
that: the degeneracy distribution $C(N+1,m)$, for small
values of $\varepsilon$, and
$m$, the number of oscillators that do fire having
 already received a pulse, for higher
values of $\varepsilon$. According to this, the different
 probabilities of selecting
pattern $m$ start being distributed following the
 degeneracy distribution $C(N+1,m)$,
and, as $\varepsilon$ decreases, these probabilities diminish
 in a hierarchical way: the
larger the value of $m$, the sooner its selection
probability is going to decrease, so
that only patterns with smaller m will survive for
 higher values of $\varepsilon$.
Moreover, some of the structures disappear, at the
 different values of
$\varepsilon_m^*$, during this process. We have found
out an approximation formula for
$p_m^{N+1}(\varepsilon)$ which takes into account all
 these mechanisms and gives us a
quantitative estimation of the different selection
 probabilities for small
$\varepsilon$.

The estimation of the volume of the basin of
 attraction of each spatio-temporal pattern
$m$ also gives us an idea of the stability of
 the different structures with respect to
additive noise fluctuations (for instance, we
 can add some random quantity $\eta$ to all
phases after each firing event or a continuous-time
 $\eta(t)$ in the driving).
Simulations of arrays of noisy pulse coupled oscillators
showed that our most probable
patterns were also the most stable\cite{PhysD}. The present
 paper only concerns
spatio-temporal pattern formation in a ring of
oscillators, nevertheless, all results
are trivially generalized to bidirectional couplings.
Although the question of what
happens when dealing with higher dimension lattices
 remains opened, some simulations
results in 2d \cite{PhysD} showed that almost all
 realizations lead to a chessboard
pattern in analogy with our results in the ring.
That makes us believe we have caught
the basic features of the problem in our 1d model.

\section*{Acknowledgments}

The authors are indebted to C.J.P\'{e}rez and A.Arenas
 for very fruitful discussions.
They also acknowledge extremely constructive suggestions
from an anonymous referee. This
work has been supported by DGICYT of the Spanish
Government through grant PB96-0168 and
EU TMR Grant ERBFMRXCT980183. X.G. also acknowledges
 financial support from the
Generalitat de Catalunya.

\end{document}